\documentclass{aastex}
\usepackage{spr-astr-addons}

\begin{document}

\title{Extensive photometry of the intermediate polar V2069~Cyg (RX J2123.7+4217)}
\shorttitle{Photometry of IP V2069~Cyg}
\shortauthors{Kozhevnikov}

\author{V. P. Kozhevnikov}
\affil{Astronomical Observatory, Ural Federal University, Lenin Av. 51, Ekaterinburg 
620083, Russia e-mail: valery.kozhevnikov@urfu.ru}

\begin{abstract} 
To obtain the spin period of the white dwarf in the intermediate polar V2069~Cyg with high precision, we fulfilled extensive photometry. Observations were performed within 32 nights, which have a total duration of 119 hours and cover 15 months. We found the spin period of the white dwarf, which is equal to $743.406\,50\pm0.000\,48$~s. Using our precise spin period, we derived the oscillation ephemeris with a long validity of 36 years. This ephemeris and the precise spin period can be used for future investigations of spin period changes of the white dwarf in V2069~Cyg. In addition, for the first time we detected the sideband oscillation with a period of $764.5125\pm0.0049$~s. The spin and sideband oscillations revealed unstable amplitudes both in a time-scale of days and in a time-scale of years. On average, the semi-amplitude of the spin oscillation varied from 17~mmag in 2014 to 25~mmag in 2015. The semi-amplitude of the sideband oscillation varied from 12~mmag in 2014 to an undetectable level of less than 7~mmag in 2015. In a time-scale of years, the optical spin pulse profile revealed essential changes from an asymmetric double-peaked shape in 2014 to a quasi-sinusoidal shape in 2015. Such drastic changes of the optical spin pulse profile seem untypical of most intermediate polars and, therefore, are of great interest. The pulse profile of the sideband oscillation was quasi-sinusoidal. Moreover, we note that V2069~Cyg possesses strong flickering with a peak-to-peak amplitude of 0.4--0.6~mag. 
\end{abstract}


\section{Introduction}

Intermediate polars (IPs), a sub-class of cataclysmic variables (CVs), are interacting binary stars, in which accretion occurs onto a magnetic white dwarf. The magnetic white dwarf spins non-synchronously with the orbital period of the system and therefore produces rapid coherent oscillations with the spin period. The spin oscillation can be observed both in optical light and in X-rays. In optical light, the spin period often appears together with the beat period, $1/P_{\rm beat} = 1/P_{\rm spin} - 1/P_{\rm orb}$. This oscillation is named the orbital sideband. Normally, the orbital sideband is produced due to the reprocessing of X-rays at some part of the system that rotates with the orbital period. This part can be the secondary star or hot spot in the accretion disc. In rare cases, the orbital sideband can be produced due to disc-overflow accretion. Then, the orbital sideband can be observed in X-rays \citep{wynn92}. Other orbital sidebands such as $\omega-2\Omega$ and $\omega+\Omega$, where $\omega=1/P_{\rm spin}$ and  $\Omega=1/P_{\rm orb}$,  can be additionally produced from the amplitude modulation with the orbital period  \citep{warner86}. A review of IPs is presented in \citet{patterson94}.

Due to a large white dwarf moment of inertia, short-period oscillations seen in IPs show a high degree of coherence. This high degree of coherence suggests that the spin period of the white dwarf can be measured with very high precision if the observational coverage is long. The precise knowledge of the spin period is important for several reasons. Firstly, at a practical level, a precise spin ephemeris allows us to phase new X-ray data with optical data (e.g., the IP Home Page, https://asd.gsfc.nasa.gov/Koji.Mukai/iphome/iphome.html).  Secondly, the precise spin period and oscillation ephemeris make it possible to perform an observational test of spin equilibrium from direct measurements of the spin period or from pulse-arrival time variations. This is an important task because many theoretical works assume that IPs are in spin equilibrium \citep[e.g.,][]{norton04}. This, however, is questionable, because only one IP, namely  FO Aqr, really proves the spin equilibrium due to alternating spin-up and spin-down \citep{patterson98, kruszewski98, williams03}. Other rare IPs with known spin rates reveal continuous spin-up or spin-down  \citep[see, e.g., table 1 in][]{warner96}. Such data are equally important because they allow us to understand angular momentum flows within the binary \citep{king99}. Moreover, the oscillation ephemeris allows one to see if any orbital variations were present in the pulse arrival time.

\begin{table}[t]
{\small 
\caption{Journal of the observations.}
\label{journal}
\begin{tabular}{@{}l c c}
\hline
\noalign{\smallskip}
Date  &  BJD$_{\rm TDB}$ start & Length \\
(UT) & (-245\,0000) & (h) \\
\hline
2014 Aug. 18   & 6888.249491 &  1.1  \\
2014 Aug. 22   & 6892.253747 & 4.5 \\
2014 Aug. 24   & 6894.224195 & 5.6   \\
2014 Aug. 25   & 6895.223789 & 2.4  \\
2014 Aug. 29   & 6899.214802 & 3.5  \\
2014 Aug. 30   & 6900.206901 & 4.5  \\
2014 Sep. 2   & 6903.217110 & 5.8  \\
2014 Sep. 16    & 6917.173205 & 2.7 \\
2014 Sep. 21    & 6922.187338 & 2.8 \\
2014 Oct. 21    & 6952.110870 & 6.1  \\
2014 Oct. 25   & 6956.268033 & 2.4  \\
2014 Oct. 26    & 6957.170653 & 5.7  \\
2014 Nov. 16    & 6978.073956 & 7.6 \\
2014 Nov. 23    & 6985.071579 & 2.3 \\
2014 Nov. 24    & 6986.177891 & 3.2 \\
2014 Nov. 27    & 6989.088255 & 5.2 \\
2015 Aug. 7   & 7242.279376 & 3.0 \\
2015 Aug. 8    & 7243.279249 & 3.0 \\
2015 Aug. 12    & 7247.322774 & 2.4 \\
2015 Aug. 15    & 7250.277559 & 1.8 \\
2015 Aug. 23    & 7258.313367 & 3.0 \\
2015 Sep. 5    & 7271.203188 & 3.9 \\
2015 Sep. 7   & 7273.263554 & 2.0 \\
2015 Sep. 10    & 7276.195874 & 1.9 \\
2015 Sep. 13    & 7279.201115 & 6.8 \\
2015 Sep. 14    & 7280.182376 & 6.1 \\
2015 Sep. 16    & 7282.267126 & 4.0 \\
2015 Sep. 22 & 7288.249860 & 4.3 \\
2015 Oct. 21   & 7317.236227 & 3.5 \\
2015 Nov. 12    & 7339.098108   & 3.4 \\
2015 Nov. 13    & 7340.157805 & 2.3 \\
2015 Nov. 18    & 7345.229705 & 2.6 \\
\hline
\end{tabular} }
\end{table}

\citet{motch96} identified the X-ray source RX~J2123.7+4217 with a new CV. This CV was subsequently called V2069~Cyg. Although, by virtue of the hard X-ray spectrum, \citeauthor{motch96} supposed that this CV belongs to the IP class, their brief photometric observations revealed no short-periodic oscillations typical of IPs. Only 14 years later \citet{demartino09} proved that this assumption is correct due to detection of the oscillation with a period of $743.2\pm 0.4$~s, which was observed both in X-rays and in optical light. Shortly afterwards this oscillation was independently confirmed both in X-rays ($743.2\pm0.9$ s, \citealt{butters11}; $743.1\pm0.6$~s, \citealt{bernardini12}) and in optical light ($743.38\pm0.25$~s, \citealt{nasiroglu12}). Thus, because all short periods measured by different authors are compatible with each other, the IP nature of V2069~Cyg seems undoubted. However, due to insufficient observational coverage, all these periods are of low precision and are not suitable to investigate changes of the spin period of the white dwarf in future or to phase different observations. Indeed, if we imagine an oscillation ephemeris with the spin period, which is measured by \citeauthor{nasiroglu12}, then the formal validity of this ephemeris will be only 26 days.  To measure the spin period with high precision and uncover other properties of V2069~Cyg, we performed extensive photometric observations within 32 nights, which have a total duration of 119 hours and cover 15 months. In this paper we present the results obtained from these observations.

\section{Observations} \label{observations}

In the observations of V2069~Cyg we used a multi-channel photometer with photomultiplier tubes. This photometer allows us to make brightness measurements of two stars and the sky background simultaneously. The design of the photometer and its noise analysis is presented in \citet{kozhevnikoviz}. The photometer is attached to the 70-cm telescope at Kourovka observatory, Ural Federal University. Advantages of this photometer in observations of IPs were proved in our old observations of V709 Cas where we discovered optical oscillations for the first time \citep{kozhevnikov01}. Later we incorporated a CCD guiding system into the photometer. Due to this guiding system, the photometer and telescope can operate automatically under computer control. This facilitates the obtaining of long continuous light curves. In addition, the precise automatic guiding improves the accuracy of brightness measurements. Such continuous light curves, which are obtained during a few tens of nights and are spread over a year, allow us to achieve very precise oscillation periods of IPs \citep[e.g.,][]{kozhevnikov12, kozhevnikov14}. 

Because V2069~Cyg is a faint star of 16~mag and is invisible by eye, to centre this star in the photometer diaphragm (16~arcsec), we used a nearby reference star and computer-controlled step motors of the telescope. The diaphragm for the comparison star was the same. However, to reduce the photon noise caused by the sky background, we measured the sky background through a diaphragm of 30 arcsec.  Photometric data of V2069~Cyg were obtained in white light (approximately 3000--8000~\AA). The time resolution was equal to 4 s. Although such a time resolution seems excessively short for the expected oscillation periods, it allows us to fill gaps in observations more accurately and decreases the errors of the periods.

The photometric observations of V2069~Cyg were obtained  in 2014 August--November over 16 nights and in 2015 August--November also over 16 nights. A journal of the observations is presented in Table~\ref{journal}. This table contains BJD$_{\rm TDB}$, which is the  Barycentric Julian Date in the Barycentric Dynamical Time (TDB) standard. TDB is a uniform time and, therefore, is preferential. We calculated BJD$_{\rm TDB}$ by using the online-calculator (http://astroutils.astronomy.ohio-state.edu/time/), which is described in \cite{eastman10}, and checked these calculations by using the BARYCEN routine in the 'aitlib' IDL library of the University of T\"{u}bingen (http://astro.uni-tuebingen.de/software/idl/aitlib/). One can change our BJD$_{\rm TDB}$ into BJD$_{\rm UTC}$, the Barycentric Julian Date in the Coordinated Universal Time (UTC) standard, by subtracting  67 s in 2004 and 68 s in 2015 \citep[e.g.,][]{eastman10}.

\begin{figure}[t]
\includegraphics[width=84mm]{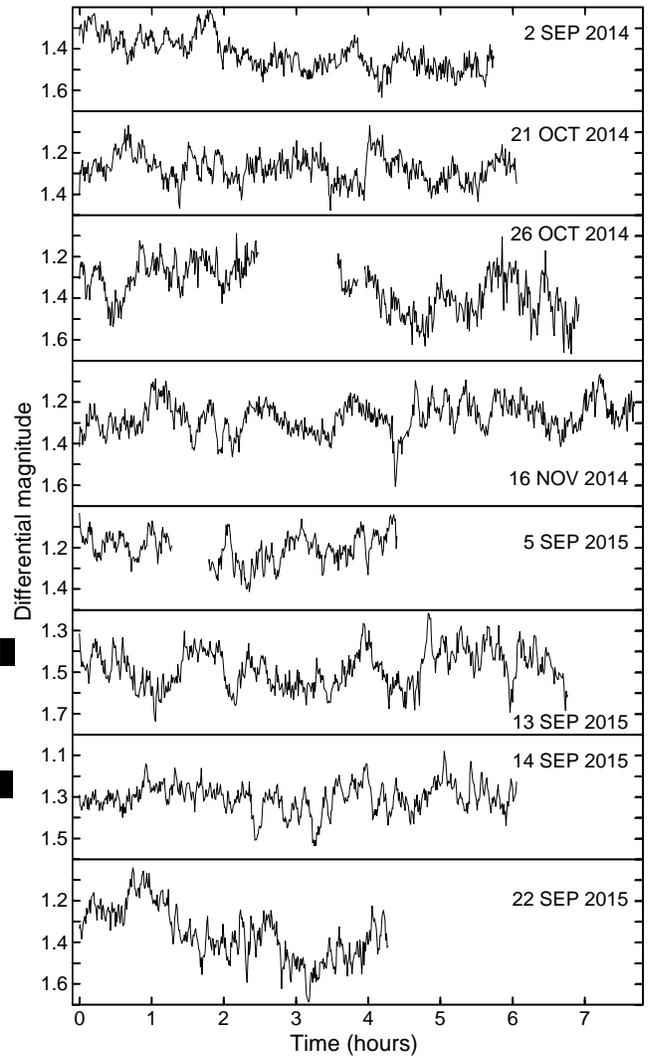}
\caption{Longest differential light curves of V2069~Cyg.}
\label{figure1}
\end{figure}

\begin{figure}[t]
\includegraphics[width=84mm]{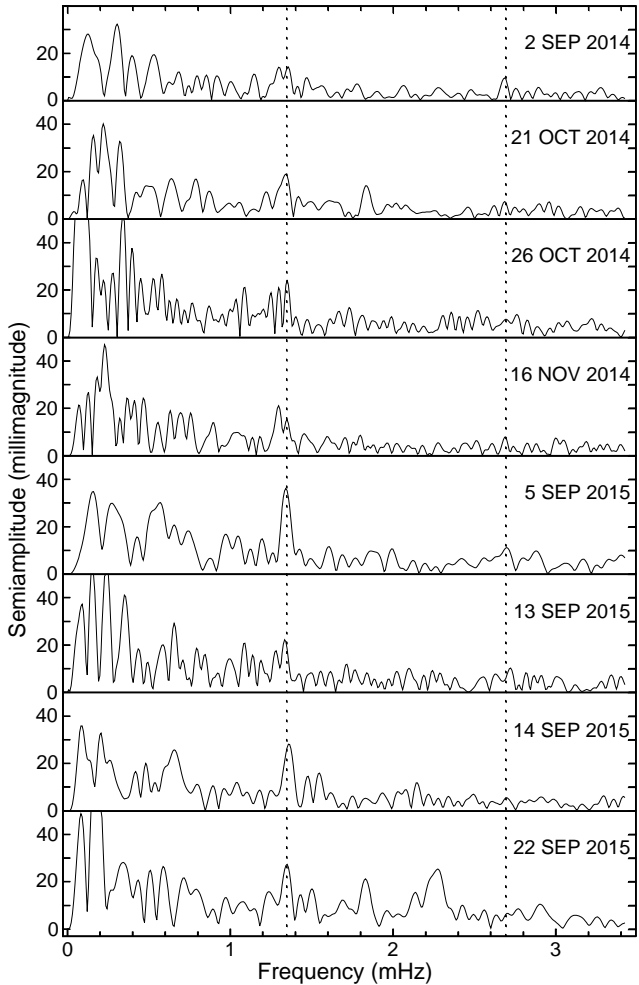}
\caption{Amplitude spectra of V2069~Cyg. The dotted lines mark the 743-s period and its first harmonic.}
\label{figure2}
\end{figure}

The comparison star is USNO-A2.0 1275-15564230. It has $B=14.9$~mag and $B-R=0.8$~mag. Its colour index is similar to the colour index of V2069~Cyg, $B-R=0.5$~mag. This reduces the influence of differential extinction. The obtained differential light curves possess the significant photon noise because of the low brightness of V2069~Cyg and high sky background. Fig.~\ref{figure1} presents four longest differential light curves obtained in 2014 and four longest differential light curves obtained in 2015, with magnitudes averaged over 40-s time intervals. The number of points in each of these light curves is in the range 351--685. According to the pulse counts of the two stars and sky background, the photon noise of these light curves (rms) is 0.02--0.04~mag.

\section{Analysis and results}

Because our multichannel photometer allows us to obtain evenly spaced data, we mainly use the classical Fourier analysis, which seems preferential in comparison with numerous methods appropriate to unevenly spaced data \citep[e.g.,][]{schwarzenberg98}. For the analysis of periodic oscillations, using a fast Fourier transform (FFT) algorithm, we calculate individual amplitude spectra, average power spectra and power spectra of the data incorporated into common time series. Before applying a FFT routine, we eliminate low-frequency trends from individual light curves by subtraction of a first- or second-order polynomial fit. This is a usual procedure in Fourier analysis and prevents discontinuity of data. This procedure does not affect high frequencies. In our previous works \citep[e.g.,][]{kozhevnikov12, kozhevnikov14} one can find details of our methods of analysis.

The longest differential light curves of V2069~Cyg presented in Fig.~\ref{figure1} show obvious flickering. Although flickering is typical of all types of CVs, the flickering power in V2069~Cyg seems noticeably stronger than the flickering power, which we observed in other IPs using the same technique. From Fig.~\ref{figure1}, we estimate the flickering peak-to-peak amplitude equal to 0.4--0.6 mag whereas V709 Cas, V515~And and V647~Cyg revealed  their flickering peak-to-peak amplitudes of less than 0.4 mag \citep[see figures 1 in][]{kozhevnikov01, kozhevnikov12,  kozhevnikov14}.

Spin oscillations of many IPs are directly visible in the light curves \citep[see, e.g., figure 1 in][]{kozhevnikov16}. In contrast, the oscillation of V2069~Cyg with a period of about 743~s, which is detected in X-rays and in optical light and obviously corresponds to the spin period \citep{bernardini12}, is inconspicuous in the light curves (Fig.~\ref{figure1}). In addition to the large flickering power, the low amplitude of the spin oscillation can be the natural reason for this invisibility. To detect this oscillation, at first we calculated amplitude spectra of individual light curves. The amplitude spectra of four longest individual light curves of 2014 and four longest individual light curves of 2015 are presented in Fig.~\ref{figure2}. As seen, these amplitude spectra clearly show peaks corresponding to the spin period only in 2015. Probably, in 2014 the spin oscillation had lesser amplitude and, therefore, was difficult to detect. One can also note that the amplitude spectra of 2014 show occasional peaks of the first harmonic of the spin period, whereas these peaks are inconspicuous in the amplitude spectra of 2015. 

\begin{figure}
\includegraphics[width=84mm]{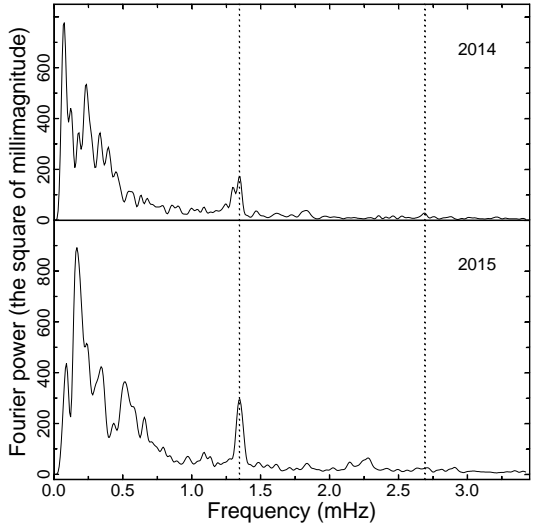}
\caption{Average power spectra derived by the weighted averaging of 10 power spectra of long individual light curves of 2014 and of 10 power spectra of long individual light curves of 2015 from V2069~Cyg. The dotted lines mark the 743-s period and its first harmonic.}
\label{figure3}
\end{figure}

\begin{figure}
\includegraphics[width=84mm]{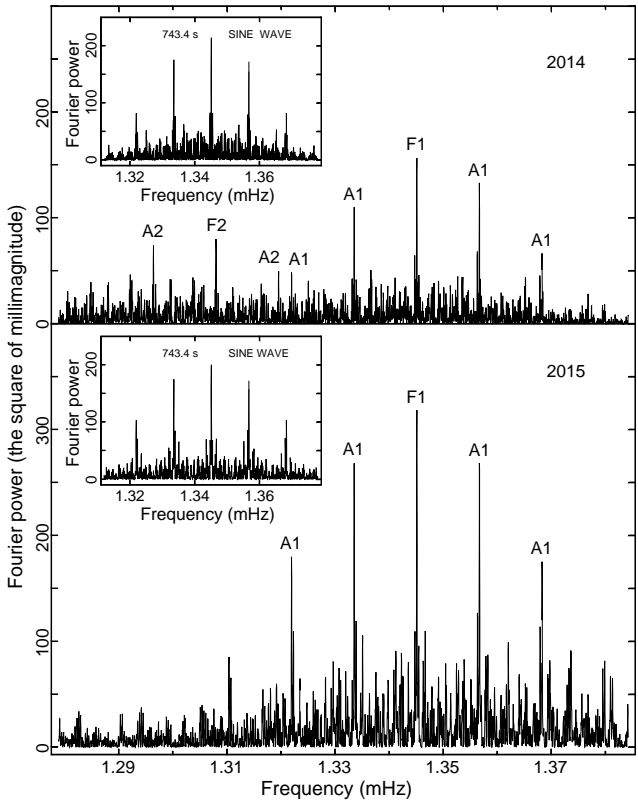}
\caption{Power spectra derived for the data of 2014 and 2015 from V2069~Cyg. They reveal a coherent oscillation with periods of $743.4060\pm0.0034$ and $743.4033\pm0.0036$~s in 2014 and 2015, respectively. In the upper frame, on the left, one can also note a sign of the sideband oscillation. The principal peaks and one-day aliases of the two oscillations are labelled with 'F1', 'F2' and 'A1', 'A2' respectively.}
\label{figure4}
\end{figure}

Fig.~\ref{figure3} presents the average power spectra calculated by the weighted averaging of the power spectra of the individual light curves, which are longer than 3~h. As seen, the spin oscillation is detectable both in the data of 2014 and in the data of 2015, but the amplitude of the spin oscillation in 2014 is somewhat lesser than in 2015. In addition, the peak of the spin oscillation in 2014 is accompanied by an additional peak on the left. Obviously, this additional peak is caused by the sideband oscillation. The intermittent presence of this peak in the amplitude spectra of 2014, which strongly changes its amplitude from night to night (Fig.~\ref{figure2}), makes the spin oscillation difficult to detect in the amplitude spectra of the individual light curves of 2014. This masking effect becomes stronger because the spin oscillation in itself changes its amplitude from night to night. This is also seen in Fig.~\ref{figure2}. Moreover, from Fig.~\ref{figure3},  we conclude that in 2015 the sideband oscillation is not detectable  at all and, therefore, does not interfere with detection of the spin oscillation. One can also note the presence of the first harmonic of the spin oscillation in 2014.  This means that the spin pulse profile is changeable from year to year, namely this profile is non-sinusoidal in 2014 and quasi-sinusoidal in 2015. This is also consistent with the occasional peaks of the first harmonic seen in the individual amplitude spectra of 2014 (Fig.~\ref{figure2}).

The average power spectra shown in Fig.~\ref{figure3} presents the overview of the periodic oscillations  in V2069~Cyg and of the behaviour of their amplitudes. These power spectra, however, do not allow us to find precise oscillation periods because of their low frequency resolution.  Therefore we analysed data incorporated into common time series, in which the gaps due to daylight and poor weather are filled with zeros. Such power spectra possess much higher frequency resolution. Fig.~\ref{figure4} shows the power spectra of two common time series containing the data of 2014 and the data of 2015 near the frequency of the spin oscillation.  As seen, the spin oscillation displays the principal peaks and one-day aliases, which are apparent from the comparison of these power spectra and the window functions shown in the insets. Most of small peaks visible in the immediate proximity of the principal peaks and one-day aliases also coincide in frequency with the corresponding small peaks of the window functions. This means that the spin oscillation is entirely coherent both during 2014 and during 2015.

Using a Gaussian function fit to upper parts of the principal peaks, we found the precise values of the spin period. These values are $743.4060\pm0.0034$ and $743.4033\pm0.0036$~s in 2014 and 2015, respectively. The errors are found according to \citet{schwarzenberg91}. These values of the spin period are compatible with each other because they differ by only $0.5\sigma$. This compatibility also confirms the conclusion made in our previous works that the errors found by the method of \citeauthor{schwarzenberg91} are true rms errors \citep{kozhevnikov12, kozhevnikov14}. The semi-amplitudes of the spin oscillation found from the power spectra shown in Fig.~\ref{figure4} are equal to 18 and 25~mmag in 2014 and 2015, respectively. These semi-amplitudes are compatible with the heights of the peaks visible in the average power spectra (Fig.~\ref{figure3}).

\begin{figure}
\includegraphics[width=84mm]{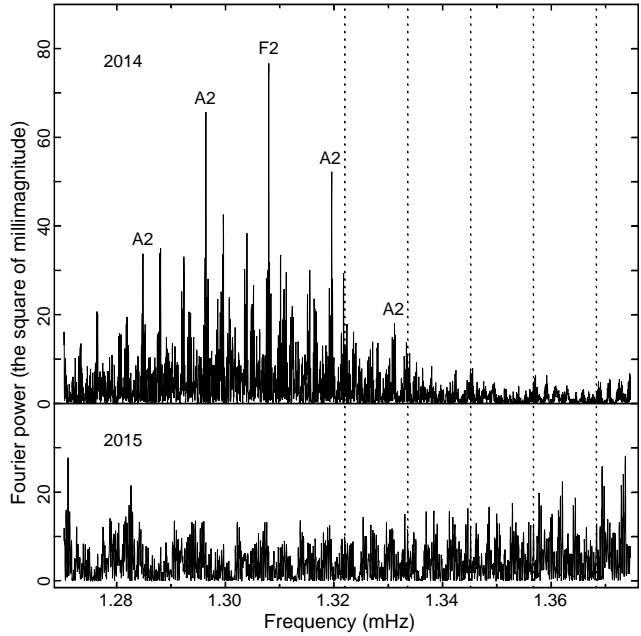}
\caption{Power spectra of the data of V2069~Cyg, from which the largest oscillation was subtracted. In the data of 2014, this subtraction allows us to detect one more coherent oscillation with a period of $764.5125\pm0.0049$~s. In the data of 2015, however, this oscillation is completely absent. The dotted lines mark the location of the principal peak of the subtracted oscillation and its one-day aliases.}
\label{figure5}
\end{figure}

As mentioned, the average power spectrum shown in the upper frame of Fig.~\ref{figure3} suggests the presence of the sideband oscillation in 2014. Obviously, additional small peaks in the upper frame of Fig.~\ref{figure4} on the left also belong to the sideband oscillation. These additional peaks, however, show no clear picture conforming to the window function because the sideband oscillation is affected by the spin oscillation, which has close frequency and higher amplitude. To remove the effect of the spin oscillation, we subtracted the spin oscillation from the data. The obtained power spectrum of the data of 2014 (the upper frame of Fig.~\ref{figure5}) clearly proves detection of  the sideband oscillation due to the presence of the principal peak and  one-day aliases conforming to the window function and showing that this oscillation is coherent. The sideband period and semi-amplitude found from this power spectrum are equal to $764.5125\pm0.0049$~s and 12~mmag, respectively.  However, the power spectrum of the pre-whitened data of 2015 (the lower frame of Fig.~\ref{figure5}) reveals the complete absence of the sideband oscillation. The semi-amplitude of the maximum noise peaks in this power spectrum is about 7~mmag. Hence, the semi-amplitude of the undetected sideband oscillation in 2015 is less than 7~mmag. The absence of signs of the sideband oscillation in the power spectrum of the common time series of 2015 conforms to the absence of this oscillation in the average power spectrum shown in the lower frame of Fig.~\ref{figure3}.

\begin{figure}
\includegraphics[width=84mm]{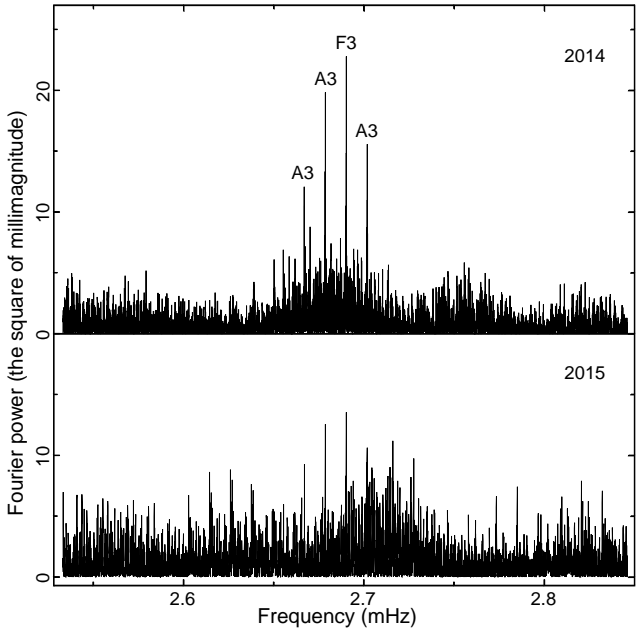}
\caption{Power spectra of the data of V2069~Cyg in the vicinity of the first harmonic of the spin oscillation. In the data of 2014, the first harmonic is clearly present, whereas, in the data of 2015, this harmonic is nearly inconspicuous among noise peaks. The principal peak and one-day aliases of the first harmonic are labelled with 'F3' and 'A3', respectively.}
\label{figure6}
\end{figure}

\begin{figure}
\includegraphics[width=84mm]{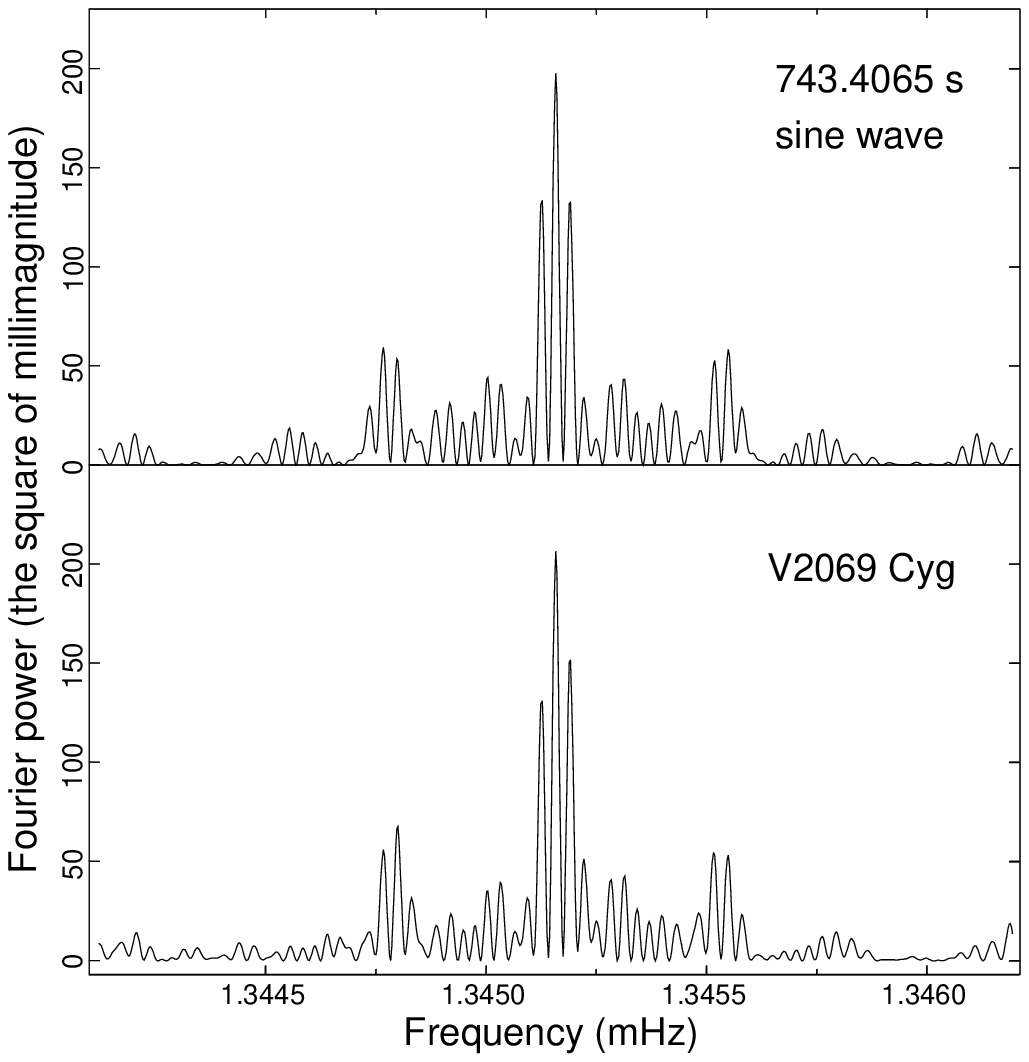}
\caption{Segment of the power spectrum calculated for all data from V2069~Cyg in the vicinity of the main oscillation. It reveals a period of $743.406\,50\pm0.000\,48$~s. The upper frame shows the window function.}
\label{figure7}
\end{figure}

Although the individual amplitude spectra (Fig.~\ref{figure2}) and average power spectra (Fig.~\ref{figure3}) suggest that the noticeable first harmonic of the spin oscillation is present only in the data of 2014, convincing evidence of this follows from the power spectra of the common time series, which are presented in Fig.~\ref{figure6}. As seen, in the data of 2014, the first harmonic of the spin oscillation reveal a distinct picture of the principal peak and one-day aliases corresponding to the window function whereas, in the data of 2015, this picture is nearly inconspicuous among noise peaks. The principal peak visible in the upper frame of Fig.~\ref{figure6} corresponds to a period of  $371.7032\pm0.0015$~s, which strictly coincides with the first harmonic of the spin oscillation. The semi-amplitude of this harmonic is equal to 7~mmag. The maximum semi-amplitude of the noise peaks visible in the lower frame of Fig.~\ref{figure6} is equal to  5~mmag.

Fig.~\ref{figure7} presents the segment of the power spectrum of the common time series, which contains both the data of 2014 and the data of 2015, in the vicinity of the spin oscillation. Obviously, this power spectrum gives the most precise spin period due to the highest frequency resolution. The period and semi-amplitude of the spin oscillation found from this power spectrum are equal to $743.406\,50\pm0.000\,48$~s and 20~mmag, respectively. As seen in Fig.~\ref{figure7}, the difference of heights of the principal peak and nearest aliases, which are caused by the large gap between the data of 2014 and 2015, is small. None the less, due to the quite high signal-to-noise ratios for the spin oscillation in the power spectra, the aliasing problem is absent. This is evident from the comparison of the spin period derived from all data and the spin periods obtained from the data 2014 and from the data of 2015 taken separately. Indeed, the deviations of the periods are less than $0.9\sigma$ when we regard the largest peak as the principal peak. However, if we suppose that the nearest alias is the true principal peak, then the deviations turn out 4--5$\sigma$. This proves the absence of the aliasing problem in the power spectrum of all data of V2069~Cyg. Unfortunately, using all data, we cannot improve the precision of the sideband period because the sideband oscillation is not detected in 2015.

Final information about the periods and amplitudes of the spin oscillation is presented in Table~\ref{table2}. The precise semi-amplitudes and their rms errors were determined from a sine wave fitted to folded light curves. Note that these semi-amplitudes are very close to the semi-amplitudes found from the power spectra. In addition, in the fourth column we give the rms errors derived by the method of \citet{schwarzenberg91}. The error of the spin period found from all data is much lower than the other errors. Therefore, we found the deviations of the other periods and expressed them in units of their rms errors. This is given in the fifth column. As seen, these deviations are not excessively small and obey a rule of $3\sigma$. This confirms our previous conclusion that the errors calculated according to \cite{schwarzenberg91} are true rms errors \citep{kozhevnikov12, kozhevnikov14}.

\begin{figure*}
\includegraphics[width=174mm]{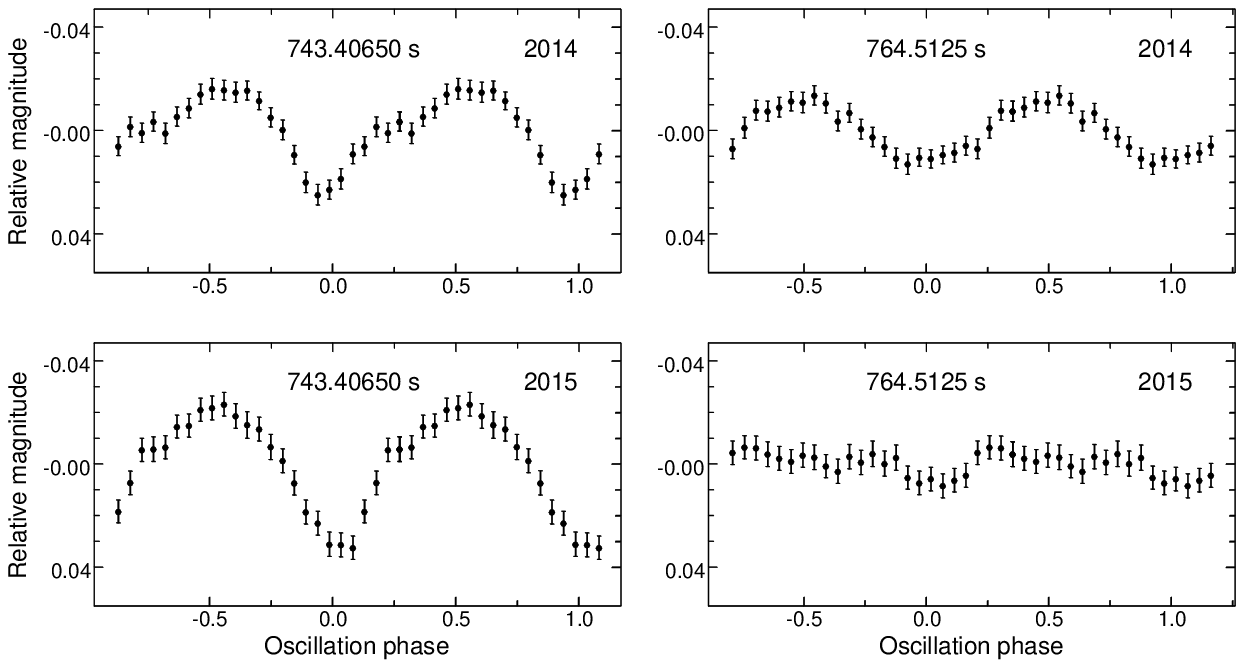}
\caption{Pulse profiles of two oscillations obtained for the data of 2014 and 2015 from V2069~Cyg. The oscillation with a period of 743.406\,50~s (on the left) reveal an unstable pulse profile, which varies from an asymmetric shape in 2014 to a quasi-sinusoidal shape in 2015. In addition, this profile has changeable amplitude and shows a small hump in phases 0.1--0.3. The oscillation with a period of 764.5125~s (on the right) has a quasi-sinusoidal pulse profile and is detected only in the data of 2014.}
\label{figure8}
\end{figure*}

\begin{table}
{\small
\caption{The values and precisions of the spin period.}
\label{table2}
\begin{tabular}{@{}l l l l l l}
\hline
\noalign{\smallskip}
Time   & Semi-amp. & Period & Error & Dev. \\
span   & (mmag)    & (s)       & $\sigma$ (s)  &     &    \\
\hline
2014 &          $17\pm1$   & 743.4060     &   0.0034   & $0.2\sigma$    \\
2015 &          $25\pm1$   & 743.4033     &   0.0036    & $0.9\sigma$    \\
Total  & $20\pm1$ & 743.406\,50   &  0.000\,48  & -- \\
\hline
\end{tabular} }
\end{table}

\begin{figure}
\includegraphics[width=84mm]{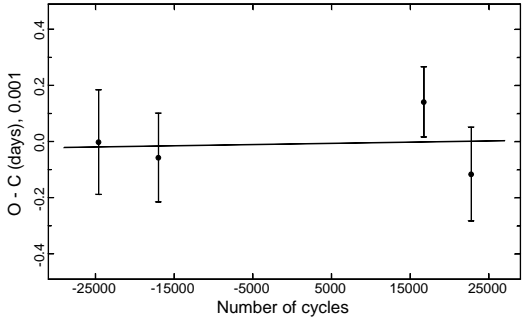}
\caption{(O--C) diagram for all data from V2069~Cyg, which are subdivided into four groups and folded with a period of $743.406\,50$~s.}

\label{figure9}

\end{figure}

\begin{figure}
\includegraphics[width=84mm]{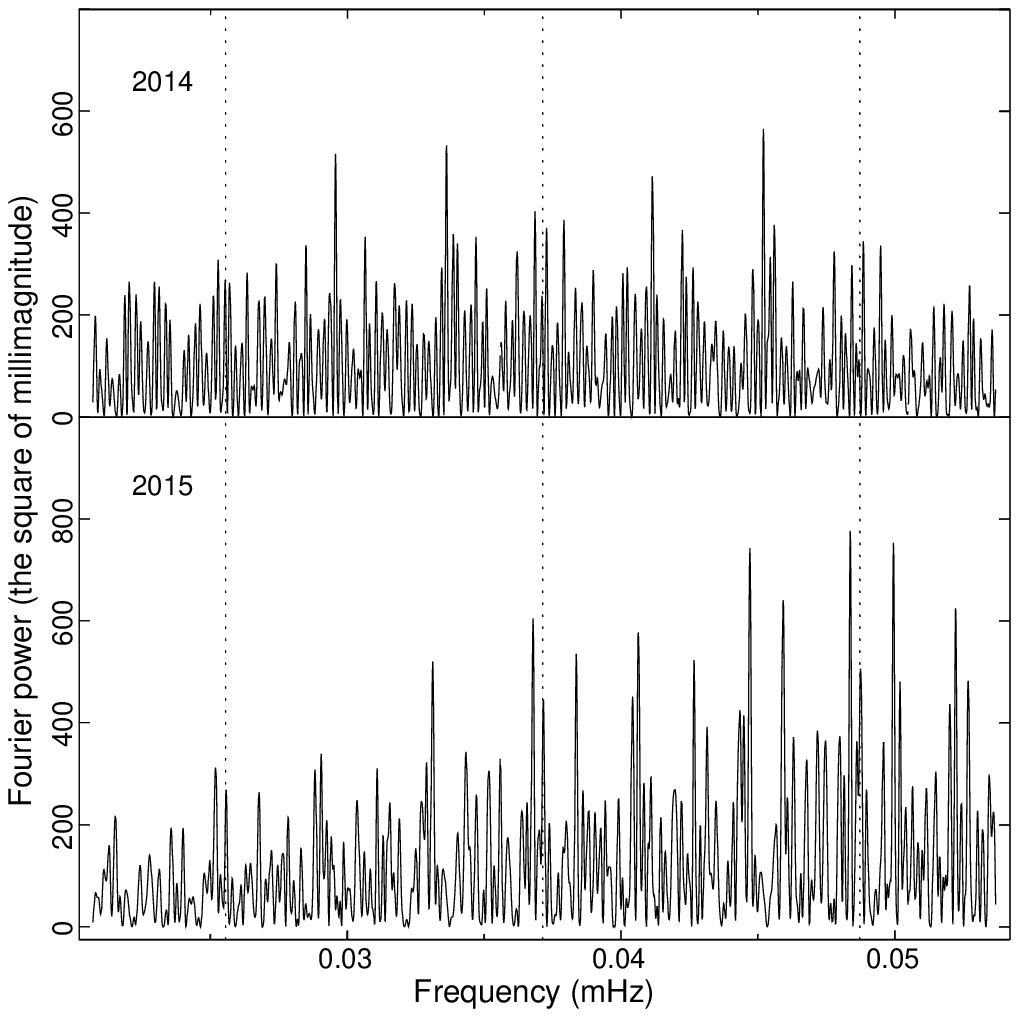}
\caption{Low-frequency parts of the power spectra for the data of 2014 and 2015 from V2069~Cyg in the frequency range of the expected orbital variability of V2069~Cyg. The dotted lines mark the expected orbital period and its nearest one-day aliases.}
\label{figure10}	
\end{figure}

Knowing the precise periods of the observed oscillations, we can obtain the average pulse profiles from the folded light curves. However, because two observed oscillations have close periods and comparable amplitudes and, in addition, the spin oscillation possesses the noticeable first harmonic, these oscillations can affect each other. To find out what kind of pre-whitening of the data is necessary to obtain unaffected pulse profiles, we performed numerical experiments with artificial time series. These three time series consisted of sine waves with the period of the fundamental harmonic of the spin oscillation, with the period of the first harmonic of the spin oscillation, with the sideband period and with the gaps according to the observations in 2014. We learned that the time series containing the sine wave with the period of the fundamental harmonic of the spin oscillation and folding with the sideband period shows a roughly sinusoidal pulse profile with amplitude of 7 per cent of the amplitude of the initial sine wave. The time series containing the sine wave with the sideband period and folding with the spin period shows the same result. The time series containing the sine wave with the period of the first harmonic of the spin oscillation and folding with the sideband period shows a double-humped pulse profile with amplitude of 3 per cent of the amplitude of the initial sine wave.

From the comparison of the amplitudes of the real oscillations and the amplitudes of the folded artificial time series we concluded that only the data, which are folded with the sideband period, require pre-whitening with the fundamental harmonic of the spin oscillation. Indeed, the semi-amplitude of the spin oscillation in 2014 is 17~mmag and, depending on the phase, can give an addition to the pulse profile of the sideband oscillation, which can reach $\pm1.2$~mmag. This amounts 10 per cent of the sideband semi-amplitude and can be appreciable against noise.  The first harmonic of the spin oscillation cannot affect the sideband pulse profile because the semi-amplitude of the first harmonic is 7~mmag and can give only a small addition of $\pm0.2$~mmag to the sideband pulse profile. Also the sideband oscillation cannot noticeably affect the spin oscillation because it can give an addition of $\pm0.8$~mmag, which is only 5 per cent of the semi-amplitude of the spin oscillation. Moreover, as seen in Fig.~\ref{figure6}, the sideband oscillation has no high-frequency harmonic and, therefore, cannot change a characteristic shape of the spin pulse profile.

Fig.~\ref{figure8} presents the light curves of V2069~Cyg folded with the spin and sideband periods. In the cases of the sideband period, the light curves were pre-whitened with the fundamental harmonic of the spin oscillation. Other kinds of pre-whitening are not necessary (see text above).  As seen, the spin oscillation (on the left) reveals an unstable pulse profile, which varies from an asymmetric shape in 2014 to a quasi-sinusoidal shape in 2015. In addition, the spin pulse profile in 2014 shows a small remarkable hump in phases 0.1--0.3. Although a weak sign of this hump is visible in 2015, we cannot consider it statistically significant. Indeed, in 2015 this hump consists of one point, which deviates from the smooth profile only by 1.5$\sigma$, whereas in 2014 this hump consists of five consecutive points, which deviate from the smooth profile by 1--2$\sigma$. Therefore, we can characterise the spin pulse profile as double-humped in 2014 and as quasi-sinusoidal in 2015.  Such characterisation is consistent with the presence of the first harmonic of the spin oscillation in the amplitude  and power spectra of 2014 and with the absence of this harmonic in the amplitude and power spectra of 2015 (Figs.~\ref{figure2},~\ref{figure3} and~\ref{figure6}).  Moreover, the spin pulse profile has changeable amplitude, which is also consistent with the oscillation amplitudes obtained from the power spectra.  The sideband oscillation (on the right) has a quasi-sinusoidal pulse profile in 2014 and is unseen in 2015. This also conforms to the amplitude and power spectra.

The high precision of the spin period makes it possible to derive an oscillation ephemeris with a long validity. To derive this ephemeris, in addition to the spin period we need in the oscillation phase. Obviously, due to a rather large noise level, the individual light curves do not allow us to find oscillation phases directly. Moreover, phases of the spin oscillation obtained from  individual light curves turn out to be shifted due to influence of  the sideband oscillation when the length of these light curves are not equal to the orbital period \citep[e.g.,][]{warner86}. Therefore, we found the time of maximum from the folding of all data, in which the effect of the sideband oscillation is inconspicuous. This time was referred to the middle of the observations. In addition, we used the data subdivided into four groups (see Table~\ref{table3}) for verification. To find the times of maxima, we might apply a Gaussian function fitted to upper parts of the maxima visible in the folded light curves.  However, as seen in Fig.~\ref{figure8}, the  spin pulse profile is changeable and, therefore,  such a method can introduce a systematic error depending on the pulse profile.  Therefore we preferred to find times of maxima by using a sine wave fit. Comparing these two methods, we found out that the corresponding times of maxima are not very different and consistent with each other within an accuracy of 13 per cent.  Finally, for the spin oscillation, we obtained the following ephemeris:

{\small
\begin{equation}
BJD_{\rm TDB} (\rm max)= 245\,7116.799\,10(7)+0.008\,604\,242(6) {\it E}.
\end{equation} }

\begin{table}
\scriptsize
\caption{Verification of the spin ephemeris.}
\label{table3}
\begin{tabular}{@{}l l l c}
\hline
\noalign{\smallskip}
Time         & BJD$_{\rm TDB}$(max)             & N. of    & O--C $\times 10^{3}$     \\
span        & (-245\,0000)                               & cycles  & (days)    \\
\noalign{\smallskip}
\hline
2014  Aug. 18--Sep.  21     & 6905.28102(11)    & --24583     &  0.00(19)   \\ 
2014 Oct. 21--Nov. 27       & 6970.70762(11)    & --16979     &  --0.06(16)   \\
2015 Aug. 7--Sep. 13           & 7260.88587(5)     & +16746     &  +0.14(13)     \\
2015 Sep. 14--Nov. 18    & 7312.76059(9)     & +22775    &  --0.12(17)   \\
\hline
\end{tabular}
\end{table}

Using this ephemeris, we obtained the (O--C) values and numbers of the oscillation cycles for the four groups of data and presented them in Table~\ref{table3}. The (O--C) values (Fig.~\ref{figure9}) reveal no significant slope and displacement along the vertical axis. Indeed, they obey the relation: ${\rm (O-C)}= - 0.000\,008(67) - 0.000\,000\,0004(33) {\it E}$. Because all quantities in this relation are less than their rms errors, the ephemeris demands no correction. From the rms error of the spin period, we found that the formal validity of this ephemeris is equal to 36 years (a $1\sigma$ confidence level).  Although this formal validity seems large, it is noticeably less than the formal validities of the spin ephemerides, which we obtained for V455 And, V647 Aur and MU Cam (85--100 years) by applying roughly the same observational coverage \citep{kozhevnikov12, kozhevnikov14, kozhevnikov16}. The reason consists in noticeably less amplitude of the spin oscillation in V2069~Cyg. Obviously, the higher relative noise level in the power spectra results in lower precision of the spin period.

The suggestion that the oscillation with a period of $764.5125\pm0.0049$~s is the orbital sideband seems quite obvious. None the less, we can check this suggestion by using the orbital period found from spectroscopic observations, $P_{\rm orb} = 7.480\,39\pm0.000\,05$~h \citep{thorstensen01}. Indeed, the orbital period calculated from the sideband and spin periods detected by us is equal to $7.4801\pm0.0017$~h. This period coincides with the spectroscopic period with high precision, where the difference is less than $0.2\sigma$. Thus, we have no doubt that the 764.5125~s period is the orbital sideband. Unfortunately, using two periods found by us, we cannot define $P_{\rm orb}$ more precisely than it was obtained by \citet{thorstensen01} because the sideband period is not detected in 2015. As mentioned, this results in much lower precision of the sideband period in comparison with the precision of the spin period.

The orbital variability of a CV can be difficult to find due to rising of the noise level at low frequencies. This increased noise level is caused by random changes of the star brightness and flickering. In the case of V2069~Cyg, this difficulty strengthens due to the unusually long orbital period. As mentioned, to detect and analyse the high-frequency oscillations, we removed the low-frequency trends from the light curves by subtraction of a first- or second-order polynomial fit. However, to search for the orbital variability, this procedure cannot be applied because most individual light curves are shorter than the long orbital period of V2069~Cyg, and low frequencies corresponding to the orbital variability will be removed. Therefore, to search for the orbital variability, we removed only nightly averages from the individual light curves. To make sure that the orbital variability of V2069~Cyg was not artificially removed due to subtraction of the nightly averages, we performed numerical experiments with artificial time series. We found out that, at least in 2014, where the individual light curves are mostly longer, this subtraction diminishes the amplitude of the orbital variability only by 20 per cent. The low-frequency parts of the obtained power spectra are shown in Fig.~\ref{figure10}. As seen, they reveal no signs of the orbital period. The semi-amplitude of the maximum noise peaks in the upper part of Fig.~\ref{figure10} is roughly 30~mmag. Therefore, we might detect the orbital variability if it had the same or larger semi-amplitude. Hence, the undetectability of the orbital period results from low orbital inclination, which must be less than $50^\circ$ \citep[e.g.,][]{ladous94}.

\section{Discussion}
We obtained photometric observations of V2069~Cyg with a total duration of 119~h in 2014 and 2015. Analysing these extensive data, we clearly detected two coherent oscillations with periods of $743.406\,50\pm0.000\,48$~s and of $764.5125\pm0.0049$~s. The shorter period is consistent with the X-ray periods found by \citet{demartino09}, \citet{butters11} and \citet{bernardini12}. \citeauthor{bernardini12} argued that V2069~Cyg is a pure disc accretor. Hence, these X-ray periods and our shorter period conform to the spin period of the white dwarf. Although this spin period was already detected in optical light \citep{nasiroglu12}, our measurement of the spin period should be considered as a new result because its precision is three orders of magnitude higher than the precisions of all other measurements of the spin period in  V2069~Cyg. The longer period detected by us is the orbital sideband because it conforms to the formula: $1/P_{\rm beat} = 1/P_{\rm spin} - 1/P_{\rm orb}$, where $P_{\rm orb}$ is the orbital period found by \citet{thorstensen01}. We detected the sideband oscillation in V2069~Cyg for the first time.

The semi-amplitude of the spin oscillation detected by us is quite low and changeable both in a time-scale of days and in a time-scale of years. On average, it is equal to 20~mmag.  In addition, we noticed that V2069~Cyg possesses flickering, which is noticeably stronger than flickering, which we observed in other IPs. Therefore, the spin oscillation is difficult to recognise directly in the light curve or to detect in the power spectrum when the light curve is short. This explains why \citet{motch96} could not find this oscillation in their first photometric observations in 1992. The amplitude of the sideband oscillation is even less than the amplitude of the spin oscillation and is also changeable. We could detect the sideband oscillation only in 2014 when its semi-amplitude was equal to 12~mmag. In 2015 the sideband oscillation completely disappeared.

In addition to changeable amplitude, the spin pulse of V2069~Cyg reveals changes of its profile. This profile varies from an asymmetric shape in 2014 to a quasi-sinusoidal shape in 2015. Moreover, the spin pulse profile observed in 2014 shows the additional small hump in phases 0.1--0.3 (the left part of Fig.~\ref{figure8}). Examining figure 6 in \citet{nasiroglu12}, we found out that in 2009 the spin pulse profile of V2069~Cyg was also double-peaked, where, in contrast with the spin pulse profile observed in 2014, two humps were equally pronounced. The reality of this pulse profile is confirmed by the power spectrum presented in figure 4 in \citeauthor{nasiroglu12}, which reveals the very strong first harmonic of the spin oscillation. Thus, between 2009 and 2015 the optical spin pulse profile of V2069~Cyg extremely varied from a pronounced double-peaked shape to a quasi-sinusoidal shape. Such drastic changes of the optical spin pulse profile seem very interesting and uncommon among other IPs.

To account for origin of optical spin pulses in IPs, three possibilities can be considered. The optical spin pulse can be produced by changes of the direct visibility of the hot pole caps, by changes of the visibility of accretion curtains located between the inner disc and the white dwarf and through reprocessing of X-rays in the axisymmetric parts of the accretion disc \citep{hellier95}. All three possibilities cannot consistently explain drastic changes of the spin pulse profile of V2069~Cyg. Indeed, in the first case, we must suppose temporary invisibility of one of the poles to the observer when the double-peaked pulse profile turns into the quasi-sinusoidal pulse profile. However, the geometry of the system cannot change and thus hide one of the poles. In the second case, depending on the sizes and shapes of the accretion curtains, both single-peaked, roughly sinusoidal and double-peaked spin pulse profiles can be generated, and these sizes and shapes depend on the magnetic field strength \citep{norton99}. Again, to account for drastic changes of the spin pulse profile, we must suppose significant changes of the magnetic field strength, which seem impossible. The third possibility seems less contradictive because changes of the spin pulse profile can originate from changes of the accretion disc structure, which seem probable. However, the reprocessing of X-rays in the axisymmetric parts of the disc demands a sufficient degree of asymmetry, e.g., between the front and the back of the disc and can happen only in a highly inclined system, which shows eclipses (e.g., DQ Her: \citealt{petterson80}; \citealt{patterson83}). Although we observed no eclipses in V2069~Cyg, none the less, we can suppose that the orbital inclination is yet sufficient to produce the spin pulse from reprocessing of X-rays in the axisymmetric parts of the disc.

Two ways are conceivable to explain origin of the optical orbital sideband in IPs. The first way consists in the reprocessing of X-rays at some structure of the system that rotates with the orbital period. The second way consists in alternation of the accretion flow between two poles of the white dwarf with the sideband frequency. The second way seems inappropriate to V2069~Cyg because this process happens in the cases stream-fed and disc-overflow accretion and because \citet{bernardini12} argued that V2069~Cyg is a pure disc accretor. Pulse profiles and amplitudes of the sideband oscillation often show significant variability.  According to the first way, which is the canonical interpretation of the optical orbital sideband, reasons for such variability can consist in changes of the structure of the accretion disc, asymmetric parts of which reprocess X-rays with the sideband frequency. This variability, however, is not accompanied by noticeable changes of the star brightness \citep{woerd84}. Then, we can account for the disappearance of the orbital sideband in 2015 by the diminishing of its amplitude to undetectable level due to changes of the structure of the accretion disc.  This explanation is also consistent with the simultaneous change of the pulse profile of the spin oscillation in V2069~Cyg because changes of the asymmetric parts of the disc can be accompanied by changes of its axisymmetric parts.

Our precise spin period of V2069~Cyg makes it possible to investigate its behaviour in future observations. This can be made by using our oscillation ephemeris, which has a formal validity of 36 years, and pulse-arrival times. Then, the alternating increase and decrease of (O--C) values can indicate spin-up and spin-down and thus indicate spin equilibrium. Of course, this is a difficult task, which demands to perform annual observations during a decade or more (see, e.g., figure 2 in \citealt{patterson98}). In addition, one can use direct measurements of the spin period obtained a decade later.  This way seems less laborious. As seen in Table~\ref{table2}, photometric observations, which are obtained during roughly 50 hours and cover a few months, give an rms error of the period of about 0.004~s. Then, performing observations during one observing season ten years later, one can detect a spin period change of 0.02~s with a 5$\sigma$ confidence level. This period change corresponds to ${\rm d}P/{\rm d}t=6 \times 10^{-11}$.  As seen in table 1 in \citet{warner96}, such a detection threshold is close to the ${\rm d}P/{\rm d}t$ measured in most IPs and, therefore, seems insufficient. Performing observations during two observing seasons ten years later, one can achieve a detection threshold of 6 times less. Probably, large ${\rm d}P/{\rm d}t $ averaged over a sufficiently large time span can imply the absence of spin equilibrium.  

\section{Conclusions}

We obtained extensive photometric observations of V2069~Cyg over 32 nights with a total duration of 119~h in 2014 and 2015. Performing comprehensive analysis of these data, we obtained the following results: 
\begin{enumerate}
\item Due to the large observational coverage, we measured the spin period of the white dwarf with high precision. The spin period is equal to $743.406\,50\pm0.000\,48$~s. 
\item The semi-amplitude of the spin oscillation was unstable both in a time-scale of days and in a time-scale of years.  On average, it varied from 17~mmag in 2014 to 25~mmag in 2015.
\item During our observations the spin pulse profile revealed strong changes. In 2014 the spin pulse profile showed an asymmetric double-peaked shape whereas in 2015 it became quasi-sinusoidal. Such drastic changes of the optical spin pulse profile seem untypical of most IPs and, therefore, are very interesting.
\item For the first time we detected the sideband oscillation of V2069~Cyg with a period of $764.5125\pm0.0049$~s.
\item The semi-amplitude of the sideband oscillation was also unstable. On average, it varied from 12~mmag in 2014 to an undetectable level of less than 7~mmag in 2015.
\item The pulse profile of the sideband oscillation was quasi-sinusoidal.
\item The high precision of the spin period allowed us to obtain the oscillation ephemeris with a formal validity of 36 years. This ephemeris and the precise spin period can be used for future investigations of spin period changes of the white dwarf in V2069~Cyg.
\item We note that V2069~Cyg possesses strong flickering with a peak-to-peak amplitude of 0.4--0.6~mag.
\end{enumerate}

\section*{Acknowledgments}

This work was supported in part by the Ministry of Education and Science (the basic part of the State assignment, RK N\textsuperscript{\underline{o}}  AAAA-A17-117030310283-7) and by the Act N\textsuperscript{\underline{o}}  211 of the Government of the Russian Federation, agreement N\textsuperscript{\underline{o}}  02.A03.21.0006. This research has made use of the SIMBAD database, operated at CDS, Strasbourg, France. This research also made use of the NASA Astrophysics Data System (ADS).

\vspace{1.0cm}
Fig. 1.  Longest differential light curves of V2069~Cyg.

\vspace{0.4cm}
Fig. 2.  Amplitude spectra of V2069~Cyg. The dotted lines mark the 743-s period and its first harmonic.

\vspace{0.4cm}
Fig. 3.  Average power spectra derived by the weighted averaging of 10 power spectra of long individual light curves of 2014 and of 10 power spectra of long individual light curves of 2015 from V2069~Cyg. The dotted lines mark the 743-s period and its first harmonic.

\vspace{0.4cm}
Fig. 4.  Power spectra derived for the data of 2014 and 2015 from V2069~Cyg. They reveal a coherent oscillation with periods of $743.4060\pm0.0034$ and $743.4033\pm0.0036$~s in 2014 and 2015, respectively. In the upper frame, on the left, one can also note a sign of the sideband oscillation. The principal peaks and one-day aliases of the two oscillations are labelled with 'F1', 'F2' and 'A1', 'A2' respectively.

\vspace{0.4cm}
Fig. 5.  Power spectra of the data of V2069~Cyg, from which the largest oscillation was subtracted. In the data of 2014, this subtraction allows us to detect one more coherent oscillation with a period of $764.5125\pm0.0049$~s. In the data of 2015, however, this oscillation is completely absent. The dotted lines mark the location of the principal peak of the subtracted oscillation and its one-day aliases.

\vspace{0.4cm}
Fig. 6.  Power spectra of the data of V2069~Cyg in the vicinity of the first harmonic of the spin oscillation. In the data of 2014, the first harmonic is clearly present, whereas, in the data of 2015, this harmonic is nearly inconspicuous among noise peaks. The principal peak and one-day aliases of the first harmonic are labelled with 'F3' and 'A3', respectively.
\vspace{0.4cm}

Fig. 7.  Segment of the power spectrum calculated for all data from V2069~Cyg in the vicinity of the main oscillation. It reveals a period of $743.406\,50\pm0.000\,48$~s. The upper frame shows the window function.

\vspace{0.4cm}
Fig. 8.  Pulse profiles of two oscillations obtained for the data of 2014 and 2015 from V2069~Cyg. The oscillation with a period of 743.406\,50~s (on the left) reveal an unstable pulse profile, which varies from an asymmetric shape in 2014 to a quasi-sinusoidal shape in 2015. In addition, this profile has changeable amplitude and shows a small hump in phases 0.1--0.3. The oscillation with a period of 764.5125~s (on the right) has a quasi-sinusoidal pulse profile and is detected only in the data of 2014.

\vspace{0.4cm}
Fig. 9.  (O--C) diagram for all data from V2069~Cyg, which are subdivided into four groups and folded with a period of $743.406\,50$~s.

\vspace{0.4cm}
Fig. 10.  Low-frequency parts of the power spectra for the data of 2014 and 2015 from V2069~Cyg in the frequency range of the expected orbital variability of V2069~Cyg. The dotted lines mark the expected orbital period and its nearest one-day aliases.

\end{document}